\documentclass[a4paper,12pt,showkeys]{revtex4}
\usepackage[utf8]{inputenc}
\usepackage[T1]{fontenc}
\usepackage{lmodern}
\usepackage{epsfig}
\usepackage{array}
\usepackage{amsmath}
\usepackage{epstopdf}
\usepackage{graphicx}
\graphicspath{/}
\begin{document}

\title{PROTON EM FORM FACTORS DATA ARE IN DISAGREEMENT WITH NEW $\sigma_{tot}(e^+e^- \to p\bar p)$ MEASUREMENTS
\footnote{ The support of the Slovak Grant Agency for Sciences VEGA, grant No.2/0153/17, is acknowledged.}}

\date{\today}

\medskip

\author{Anna Z. Dubni\v ckov\'a}
\address{Department of Theoretical Physics, Comenius University, Mlynska dolina,
SK-84248 Bratislava, Slovak Republic}

\author{Stanislav Dubni\v cka}
\address{Institute of Physics, Slovak Academy of Sciences, Dubravska cesta 9,
SK-84511 Bratislava, Slovak Republic}

\begin{abstract}
   Till now almost 50 different data collections on the proton and neutron electromagnetic form factors, also on their ratios, exist. On the other hand, recently new very precise data on $\sigma_{tot}(e^+e^- \to p \bar p)$ from threshold up to cca 13.5 GeV$^2$ have been obtained by BESIII Collaboration using the initial state radiation technique with an undetected photon at the BEPcII collider. In this paper a consistency of the  $\sigma_{tot}(e^+e^- \to p \bar p)$ data with
   the proton form factors behaviors is investigated, first when the latter are obtained in the analysis of the proton and
   neutron form factors data together by the advanced nucleon electromagnetic structure Unitary and Analytic model, and then, as the neutron data are considered to be not very precise, obtained in the analysis of only the proton form factors data in spacelike and timelike regions by the
   advanced proton electromagnetic structure Unitary and Analytic model. In both cases one finds the new $\sigma_{tot}(e^+e^- \to p \bar p)$
   data to be inconsistent with the curves of $\sigma_{tot}(e^+e^- \to p \bar p)$ calculated by the proton electric and magnetic form factors behaviors found in both analyses.
\end{abstract}


\maketitle

\newpage
\section{Introduction}

   The electromagnetic (EM) structure of the nucleon (isodublet compound of the proton and neutron)
is completely described theoretically by two independent functions of one variable, the Dirac $F^N_1(t)$ and Pauli $F^N_2(t)$ form factors (FFs),
which naturally appear in a decomposition of the nucleon matrix element of the EM current $J^{EM}_\mu (0)$ as coefficients of two
linearly independent covariants constructed from the four momenta $p, p'$,
$\gamma$-matrices and Dirac bi-spinors
\begin{small}
\begin{eqnarray}
  <N|J^{EM}_\mu (0)|N>=e \bar u(p')[\gamma_\mu F^N_1(t)+\frac{i}{2m_N}
  \sigma_{\mu \nu}(p'-p)_\mu F^N_2(t)] u(p),
\end{eqnarray}
\end{small}
with $m_N$ to be the nucleon mass.

   A description of EM structure of the nucleon is even improved if mixed transformation properties of the
EM current $J^{EM}_\mu (0)$ under the rotation in the isospin space is utilized. A part of $J^{EM}_\mu (0)$ transforms as an isoscalar and its another part as the third component of isovector. The latter leads to a splitting of the proton and neutron Dirac and Pauli EM FFs to flavour-independent isoscalar and isovector parts $F^N_{1s}(t), F^N_{1v}(t), F^N_{2s}(t), F^N_{2v}(t)$ as follows
\begin{small}
\begin{eqnarray}
 F^p_1(t)=[F^N_{1s}(t)+F^N_{1v}(t)]\nonumber\\
 F^p_2(t)=[F^N_{2s}(t)+F^N_{2v}(t)]\\
 F^n_1(t)=[F^N_{1s}(t)-F^N_{1v}(t)]\nonumber\\
 F^n_2(t)=[F^N_{2s}(t)-F^N_{2v}(t)],\nonumber
\end{eqnarray}
\end{small}
whereby the sign between them is specified by the sign of the third component of the isospin
of the concrete nucleon under consideration.

   The FFs $F^N_{1s}(t), F^N_{1v}(t), F^N_{2s}(t), F^N_{2v}(t)$ are analytic in the whole complex t-plane besides cuts
on the positive real axis starting for isovector FFs at the two-pion threshold and for isoscalar FFs at three-pion threshold.

   In the paper \cite{ABDD} the advanced 9 vector-meson resonance Unitary and Analytic $(U\&A)$ model for
nucleon isoscalar and isovector Dirac and Pauli FFs has been constructed
\begin{small}
\begin{eqnarray}\label{FN1s}\nonumber
  F^N_{1s}[V(t)]=\Bigg(\frac{1-V^2}{1-V^2_N}\Bigg)^4\Bigg\{\frac{1}{2}H_{\omega''}(V)H_{\phi''}(V)\\\nonumber
  +\Bigg[H_{\phi''}(V)H_{\omega'}(V)\frac{(C^{1s}_{\phi''}-C^{1s}_{\omega'})}{(C^{1s}_{\phi''}-C^{1s}_{\omega''})}+
  H_{\omega''}(V)H_{\omega'}(V)\frac{(C^{1s}_{\omega''}-C^{1s}_{\omega'})}{(C^{1s}_{\omega''}-C^{1s}_{\phi''})}\\
  -H_{\omega''}(V)H_{\phi''}(V)\Bigg](f^{(1)}_{\omega'NN}/f_{\omega'})\nonumber\\
  +\Bigg[H_{\phi''}(V)H_{\phi'}(V)\frac{(C^{1s}_{\phi''}-C^{1s}_{\phi'})}{(C^{1s}_{\phi''}-C^{1s}_{\omega''})}+
  H_{\omega''}(V)H_{\phi'}(V)\frac{(C^{1s}_{\omega''}-C^{1s}_{\phi'})}{(C^{1s}_{\omega''}-C^{1s}_{\phi''})}\nonumber\\
  -H_{\omega''}(V)H_{\phi''}(V)\Bigg](f^{(1)}_{\phi'NN}/f_{\phi'})\\\nonumber
  +\Bigg[H_{\phi''}(V)L_{\omega}(V)\frac{(C^{1s}_{\phi''}-C^{1s}_{\omega})}{(C^{1s}_{\phi''}-C^{1s}_{\omega''})}+
  H_{\omega''}(V)L_{\omega}(V)\frac{(C^{1s}_{\omega''}-C^{1s}_{\omega})}{(C^{1s}_{\omega''}-C^{1s}_{\phi''})}\\\nonumber
  -H_{\omega''}(V)H_{\phi''}(V)\Bigg](f^{(1)}_{\omega NN}/f_{\omega})\\\nonumber
  +\Bigg[H_{\phi''}(V)L_{\phi}(V)\frac{(C^{1s}_{\phi''}-C^{1s}_{\phi})}{(C^{1s}_{\phi''}-C^{1s}_{\omega''})}+
  H_{\omega''}(V)L_{\phi}(V)\frac{(C^{1s}_{\omega''}-C^{1s}_{\phi})}{(C^{1s}_{\omega''}-C^{1s}_{\phi''})}\\
  -H_{\omega''}(V)H_{\phi''}(V)\Bigg](f^{(1)}_{\phi NN}/f_{\phi})\Bigg\}\nonumber
\end{eqnarray}
\end{small}
with 5 free parameters
\begin{small}
$(f^{(1)}_{\omega'NN}/f_{\omega'}), (f^{(1)}_{\phi'NN}/f_{\phi'}),
(f^{(1)}_{\omega NN}/f_{\omega}), (f^{(1)}_{\phi NN}/f_{\phi}),
t^{1s}_{in}$
\end{small}
\begin{small}
\begin{eqnarray}\label{FN1v}\nonumber
  F^N_{1v}[W(t)]=\Bigg(\frac{1-W^2}{1-W^2_N}\Bigg)^4\Bigg\{\frac{1}{2}L_\rho(W)L_{\rho'}(W)\\
  +\Bigg[L_{\rho'}(W)L_{\rho''}(W)\frac{(C^{1v}_{\rho'}-C^{1v}_{\rho''})}{(C^{1v}_{\rho'}-C^{1v}_\rho)}+
  L_\rho(W)L_{\rho''}(W)\frac{(C^{1v}_\rho-C^{1v}_{\rho''})}{(C^{1v}_\rho-C^{1v}_{\rho'})}\\\nonumber
  -L_\rho(W)L_{\rho'}(W)\Bigg](f^{(1)}_{\rho NN}/f_{\rho})\Bigg\}
\end{eqnarray}
\end{small}
with 2 free parameters
$(f^{(1)}_{\rho NN}/f_{\rho})$ and $t^{1v}_{in}$,
\begin{small}
\begin{eqnarray}\label{FN2s}\nonumber
  F^N_{2s}[U(t)]=\Bigg(\frac{1-U^2}{1-U^2_N}\Bigg)^6\Bigg\{\frac{1}{2}(\mu_p+\mu_n-1)H_{\omega''}(U)H_{\phi''}(U)H_{\omega'}(U)\\\nonumber
  +\Bigg[H_{\phi''}(U)H_{\omega'}(U)H_{\phi'}(U)\frac{(C^{2s}_{\phi''}-C^{2s}_{\phi'})(C^{2s}_{\omega'}-C^{2s}_{\phi'})}
  {(C^{2s}_{\phi''}-C^{2s}_{\omega''})(C^{2s}_{\omega'}-C^{2s}_{\omega''})}\\\nonumber
  +H_{\omega''}(U)H_{\omega'}(U)H_{\phi'}(U)\frac{(C^{2s}_{\omega''}-C^{2s}_{\phi'})(C^{2s}_{\omega'}-C^{2s}_{\phi'})}
  {(C^{2s}_{\omega''}-C^{2s}_{\phi''})(C^{2s}_{\omega'}-C^{2s}_{\phi''})}\\\nonumber
  +H_{\omega''}(U)H_{\phi''}(U)H_{\phi'}(U)\frac{(C^{2s}_{\omega''}-C^{2s}_{\phi'})(C^{2s}_{\phi''}-C^{2s}_{\phi'})}
  {(C^{2s}_{\omega''}-C^{2s}_{\omega'})(C^{2s}_{\phi''}-C^{2s}_{\omega'})}\\\nonumber
  -H_{\omega''}(U)H_{\phi''}(U)H_{\omega'}(U)\Bigg](f^{(2)}_{\phi'NN}/f_{\phi'})\\\nonumber
  +\Bigg[H_{\phi''}(U)H_{\omega'}(U)L_{\omega}(U)\frac{(C^{2s}_{\phi''}-C^{2s}_{\omega})(C^{2s}_{\omega'}-C^{2s}_{\omega})}
  {(C^{2s}_{\phi''}-C^{2s}_{\omega''})(C^{2s}_{\omega'}-C^{2s}_{\omega''})}\\\nonumber
  +H_{\omega''}(U)H_{\omega'}(U)L_{\omega}(U)\frac{(C^{2s}_{\omega''}-C^{2s}_{\omega})(C^{2s}_{\omega'}-C^{2s}_{\omega})}
  {(C^{2s}_{\omega''}-C^{2s}_{\phi''})(C^{2s}_{\omega'}-C^{2s}_{\phi''})}+\\
  +H_{\omega''}(U)H_{\phi''}(U)L_{\omega}(U)\frac{(C^{2s}_{\omega''}-C^{2s}_{\omega})(C^{2s}_{\phi'}-C^{2s}_{\omega})}
  {(C^{2s}_{\omega''}-C^{2s}_{\omega'})(C^{2s}_{\phi''}-C^{2s}_{\omega'})}\\\nonumber
  -H_{\omega''}(U)H_{\phi''}(U)H_{\omega'}(U)\Bigg](f^{(2)}_{\omega NN}/f_{\omega})\\\nonumber
  +\Bigg[H_{\phi''}(U)H_{\omega'}(U)L_{\phi}(U)\frac{(C^{2s}_{\phi''}-C^{2s}_{\phi})(C^{2s}_{\omega'}-C^{2s}_{\phi})}
  {(C^{2s}_{\phi''}-C^{2s}_{\omega''})(C^{2s}_{\omega'}-C^{2s}_{\omega''})}\\\nonumber
  +H_{\omega''}(U)H_{\omega'}(U)L_{\phi}(U)\frac{(C^{2s}_{\omega''}-C^{2s}_{\phi})(C^{2s}_{\omega'}-C^{2s}_{\phi})}
  {(C^{2s}_{\omega''}-C^{2s}_{\phi''})(C^{2s}_{\omega'}-C^{2s}_{\phi''})}\\\nonumber
  +H_{\omega''}(U)H_{\phi''}(U)L_{\phi}(U)\frac{(C^{2s}_{\omega''}-C^{2s}_{\phi})(C^{2s}_{\phi''}-C^{2s}_{\phi})}
  {(C^{2s}_{\omega''}-C^{2s}_{\omega'})(C^{2s}_{\phi''}-C^{2s}_{\omega'})}\\\nonumber
  -H_{\omega''}(U)H_{\phi''}(U)H_{\omega'}(U)\Bigg](f^{(2)}_{\phi NN}/f_{\phi})\Bigg\}
\end{eqnarray}
\end{small}
with 4 free parameters
$(f^{(2)}_{\phi'NN}/f_{\phi'})$, $(f^{(2)}_{\omega NN}/f_{\omega})$,
$(f^{(2)}_{\phi NN}/f_{\phi}), t^{2s}_{in}$, and
\begin{small}
\begin{eqnarray}\label{FN2v}
  F^N_{2v}[X(t)]=\Bigg(\frac{1-X^2}{1-X^2_N}\Bigg)^6\Bigg\{\frac{1}{2}(\mu_p-\mu_n-1)L_\rho(X)L_{\rho'}(X)H_{\rho''}(X)\Bigg\}
\end{eqnarray}
\end{small}
dependent on only 1 free parameter
$t^{2v}_{in}$, where

\begin{small}
\begin{eqnarray}
  L_r(V)=\frac{(V_N-V_r)(V_N-V^*_r)(V_N-1/V_r)(V_N-1/V^*_r)}{(V-V_r)(V-V^*_r)(V-1/V_r)(V-1/V^*_r)},\\\label{eq19}
  C^{1s}_r=\frac{(V_N-V_r)(V_N-V^*_r)(V_N-1/V_r)(V_N-1/V^*_r)}{-(V_r-1/V_r)(V_r-1/V^*_r)}, r=\omega, \phi \nonumber
\end{eqnarray}
\end{small}
\begin{small}
\begin{eqnarray}
  H_l(V)=\frac{(V_N-V_l)(V_N-V^*_l)(V_N+V_l)(V_N+V^*_l)}{(V-V_l)(V-V^*_l)(V+V_l)(V+V^*_l)},\\\label{eq20}
  C^{1s}_l=\frac{(V_N-V_l)(V_N-V^*_l)(V_N+V_l)(V_N+V^*_l)}{-(V_l-1/V_l)(V_l-1/V^*_l)}, l=
  \omega'', \phi'', \omega', \phi' \nonumber
\end{eqnarray}
\end{small}
\begin{small}
\begin{eqnarray}
  L_k(W)=\frac{(W_N-W_k)(W_N-W^*_k)(W_N-1/W_k)(W_N-1/W^*_k)}{(W-W_k)(W-W^*_k)(W-1/W_k)(W-1/W^*_k)},\\ \label{eq21}
  C^{1v}_k=\frac{(W_N-W_k)(W_N-W^*_k)(W_N-1/W_k)(W_N-1/W^*_k)}{-(W_k-1/W_k)(W_k-1/W^*_k)}, k=\rho'',
  \rho', \rho \nonumber
\end{eqnarray}
\end{small}
\begin{small}
\begin{eqnarray}
  L_r(U)=\frac{(U_N-U_r)(U_N-U^*_r)(U_N-1/U_r)(U_N-1/U^*_r)}{(U-U_r)(U-U^*_r)(U-1/U_r)(U-1/U^*_r)},\\ \label{eq22}
  C^{2s}_r=\frac{(U_N-U_r)(U_N-U^*_r)(U_N-1/U_r)(U_N-1/U^*_r)}{-(U_r-1/U_r)(U_r-1/U^*_r)}, r=\omega, \phi \nonumber
\end{eqnarray}
\end{small}
\begin{small}
\begin{eqnarray}
  H_l(U)=\frac{(U_N-U_l)(U_N-U^*_l)(U_N+U_l)(U_N+U^*_l)}{(U-U_l)(U-U^*_l)(U+U_l)(U+U^*_l)},\\ \label{eq23}
  C^{2s}_l=\frac{(U_N-U_l)(U_N-U^*_l)(U_N+U_l)(U_N+U^*_l)}{-(U_l-1/U_l)(U_l-1/U^*_l)}, l=
  \omega'', \phi'', \omega', \phi' \nonumber
\end{eqnarray}
\end{small}
\begin{small}
\begin{eqnarray}
  L_k(X)=\frac{(X_N-X_k)(X_N-X^*_k)(X_N-1/X_k)(X_N-1/X^*_k)}{(X-X_k)(X-X^*_k)(X-1/X_k)(X-1/X^*_k)},\\\label{eq24}
  C^{2v}_k=\frac{(X_N-X_k)(X_N-X^*_k)(X_N-1/X_k)(X_N-1/X^*_k)}{-(X_k-1/X_k)(X_k-1/X^*_k)}, k=\rho', \rho \nonumber
\end{eqnarray}
\end{small}
\begin{small}
\begin{eqnarray}
  H_{\rho''}(X)=\frac{(X_N-X_{\rho''})(X_N-X^*_{\rho''})(X_N+X_{\rho''})(X_N+X^*_{\rho''})}
  {(X-X_{\rho''})(X-X^*_{\rho''})(X+X_{\rho''})(X+X^*_{\rho''})},\\ \label{eq25}
  C^{2v}_{\rho''}=\frac{(X_N-X_{\rho''})(X_N-X^*_{\rho''})(X_N+X_{\rho''})(X_N+X^*_{\rho''})}
  {-(X_{\rho''}-1/X_{\rho''})(X_{\rho''}-1/X^*_{\rho''})},\nonumber
\end{eqnarray}
\end{small}
which involves besides the lowest branch points $t^{1s}_0=9m_\pi^2$, $t^{2s}_0=9m_\pi^2$ and $t^{1v}_0=4m_\pi^2$, $t^{2v}_0=4m_\pi^2$ also
square root branch points $t^{1s}_{in}$, $t^{2s}_{in}$, $t^{1v}_{in}$, $t^{2v}_{in}$, giving contributions of all possible higher inelastic
thresholds effectively and therefore they are left in the analysis of data as free parameters.

   This model corresponds to nowadays experimentally confirmed nine neutral vector-mesons \cite{PDG} $\rho(770),
\omega(782), \phi(1020); \rho'(1450), \omega'(1420), \phi'(1680); \rho''(1700), \omega''(1650), \phi''(2170);$ with quantum numbers of the photon and is in fact a well-matched unification of pole contributions of unstable vector mesons with cut structure in the complex plane $t$,
whereby these cuts represent so-called continua contributions generated by exchange of more than one particle in the corresponding Feynman diagrams.

    As a result in such model the shape of nucleon EM FFs is directly related to the existence of complex
conjugate pairs of unstable vector-meson poles on three unphysical sheets of the four-sheeted Riemann surface in $t$ variable
and cut contributions representing continua additions are securing FFs to be complex beyond the lowest possible thresholds on the positive real axis,
as it is required by the FFs unitarity conditions.

   Here we would like to note that the nucleon Dirac $F^N_1(t)$ and Pauli $F^N_2(t)$ FFs, as it is seen from the
previous, are very suitable for theoretical description of the nucleon EM structure. However, for an extraction of experimental information on the nucleon EM structure from measured cross sections and polarizations, the nucleon Sachs EM FFs $G^N_E(t), G^N_M(t)$ are more suitable, which appear e.g. in the total cross section of $e^+e^- \to N \bar N$ process  \cite{BPZZ}
\begin{eqnarray}\label{totcspp}
 \sigma_{tot}(e^+e^- \to N \bar N)=\frac{4 \pi \alpha^2 C \beta_N(t)}{3 t}
 [|G^N_M(t)|^2+\frac{2m_N^2}{t}|G^N_E(s)|^2]
\end{eqnarray}
with $\beta_N(t)=\sqrt{1-\frac{4 m^2_N}{t}}$ and $C$ to be the Coulomb enhancement factor, without interference term, unlike Dirac and Pauli FFs.

   The relations between the nucleon Sachs EM FFs $G^N_E(t), G^N_M(t)$ and the isoscalar and isovector
parts of the nucleon Dirac and Pauli FFs are,\\
for proton
\begin{eqnarray}\label{pEMFFs}
  G^p_E(t)=[F^N_{1s}(t)+F^N_{1v}(t)]+
  \frac{t}{4 m^2_p}[F^N_{2s}(t)+F^N_{2v}(t)]\\
  G^p_M(t)=[F^N_{1s}(t)+F^N_{1v}(t)]+[F^N_{2s}(t)+F^N_{2v}(t)]\nonumber
\end{eqnarray}
and for neutron
\begin{eqnarray}\label{nEMFFs}
  G^n_E(t)=[F^N_{1s}(t)-F^N_{1v}(t)]+
  \frac{t}{4 m^2_n}[F^N_{2s}(t)-F^N_{2v}(t)]\\
  G^n_M(t)=[F^N_{1s}(t)-F^N_{1v}(t)]+[F^N_{2s}(t)-F^N_{2v}(t)]\nonumber
\end{eqnarray}
with normalizations
\begin{eqnarray}
  G^p_E(0)= 1;\quad G^p_M(0)=\mu_p;\quad G^n_E(0)= 0;\quad G^n_M(0)=\mu_n;
\end{eqnarray}
and
\begin{eqnarray}
  F^N_{1s}(0)=F^N_{1v}(0)=\frac{1}{2};\quad F^N_{2s}(0)=\frac{1}{2}(\mu_p+\mu_n-1);\quad F^N_{2v}(0)=\frac{1}{2}(\mu_p-\mu_n-1),
\end{eqnarray}
where $\mu_N$ $N=p,n$ are the magnetic moments of the proton and neutron, respectively.

\section{Experimental information on nucleon EM FFs and its analysis}

   The experimental information on the nucleon electric $G_{Ep}(t), G_{En}(t)$ and
magnetic $G_{Mp}(t), G_{Mn}(t)$ FFs consists of the following different sets of data
\begin{itemize}
\item
  the ratio $\mu_p G^p_E(t)/G^p_M(t)$ in the space-like $(t<0)$ region
  from polarization experiments \cite{jones}-\cite{puckett}
\item
  $G^p_E(t)$ in the space-like $(t<0)$ region \cite{bernauer}
\item
  $|G^p_E(t)|$ in the time-like $(t>0)$ region; only from experiments in
  which $|G^p_E(t)|$=$|G^p_M(t)|$ is assumed and \cite{Ab1}
\item
  $G^p_Mt)$ in the space-like $(t<0)$ region
  \cite{bernauer}-\cite{andivakis}
\item
  $|G^p_M(t)|$ in the time-like $(t>0)$ region
  \cite{ablikim}-\cite{lees2}, \cite{Ab1}
\item
  $|G^p_E(t)/G^p_M(t)|$ in the time-like $(t>0)$ region
  \cite{ablikim},\cite{lees},\cite{Ab1}
\item
  $G^n_E(t)$ in the space-like $(t<0)$ region \cite{hanson}-\cite{golak}
\item
  $|G^n_E(t)|$ in the time-like $(t>0)$ region; only from experiment in
  which $|G^n_E(t)|$=$|G^n_M(t)|$ is assumed
\item
  $G^n_M(t)$ in the space-like $(t<0)$ region
  \cite{hanson},\cite{rock}-\cite{anderson}
\item
  $|G^n_M(t)|$ in the time-like $(t>0)$ region \cite{antonelli}
\item
  the ratio $\mu_n G^n_E(t)/G^n_M(t)$ in the space-like $(t<0)$ region
  from polarization experiments on the light nuclei
  \cite{plaster},\cite{riordan}.
\end{itemize}

\begin{figure}
    \includegraphics[width=0.25\textwidth]{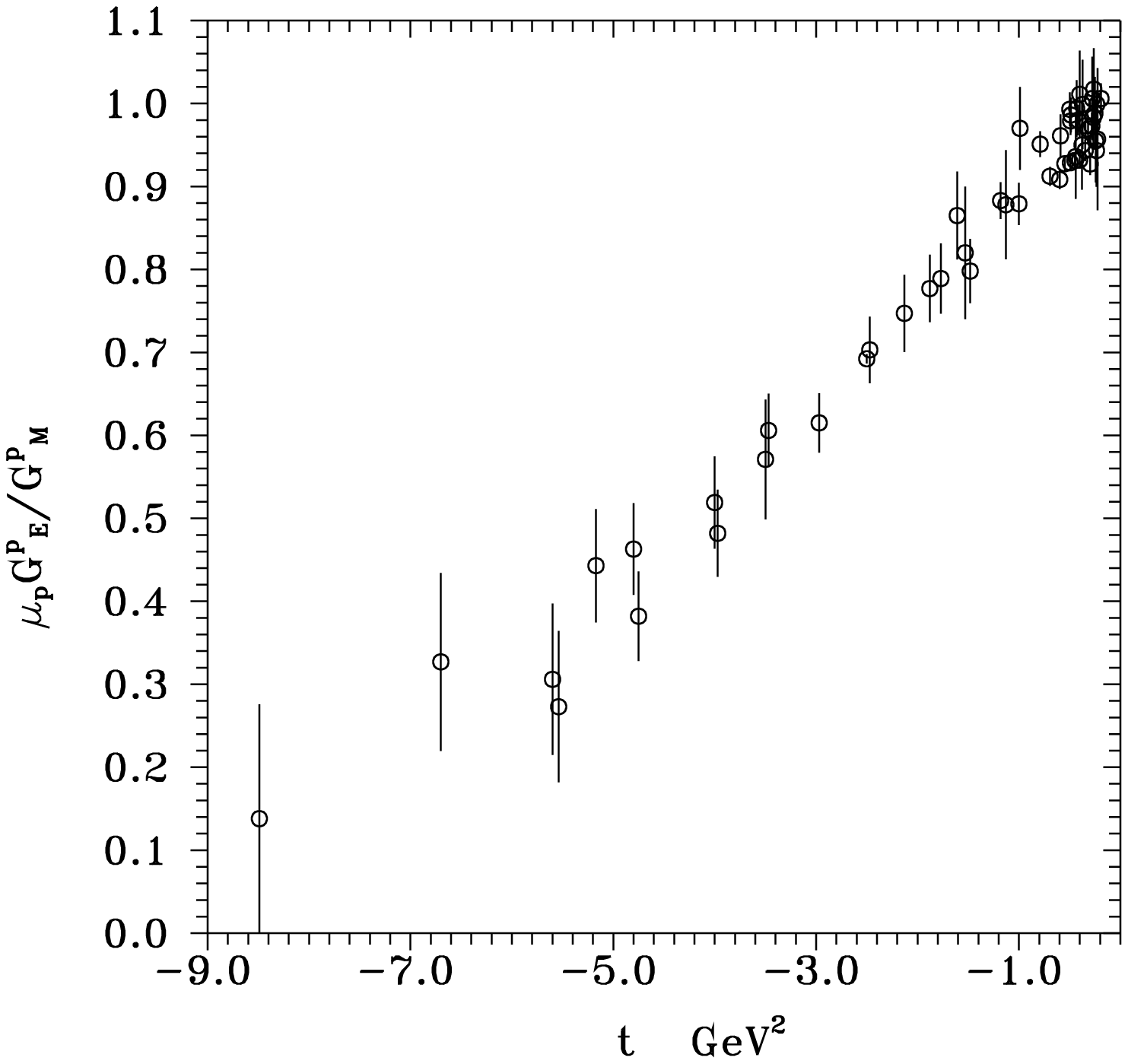}\hspace{0.3cm}
    \includegraphics[width=0.25\textwidth]{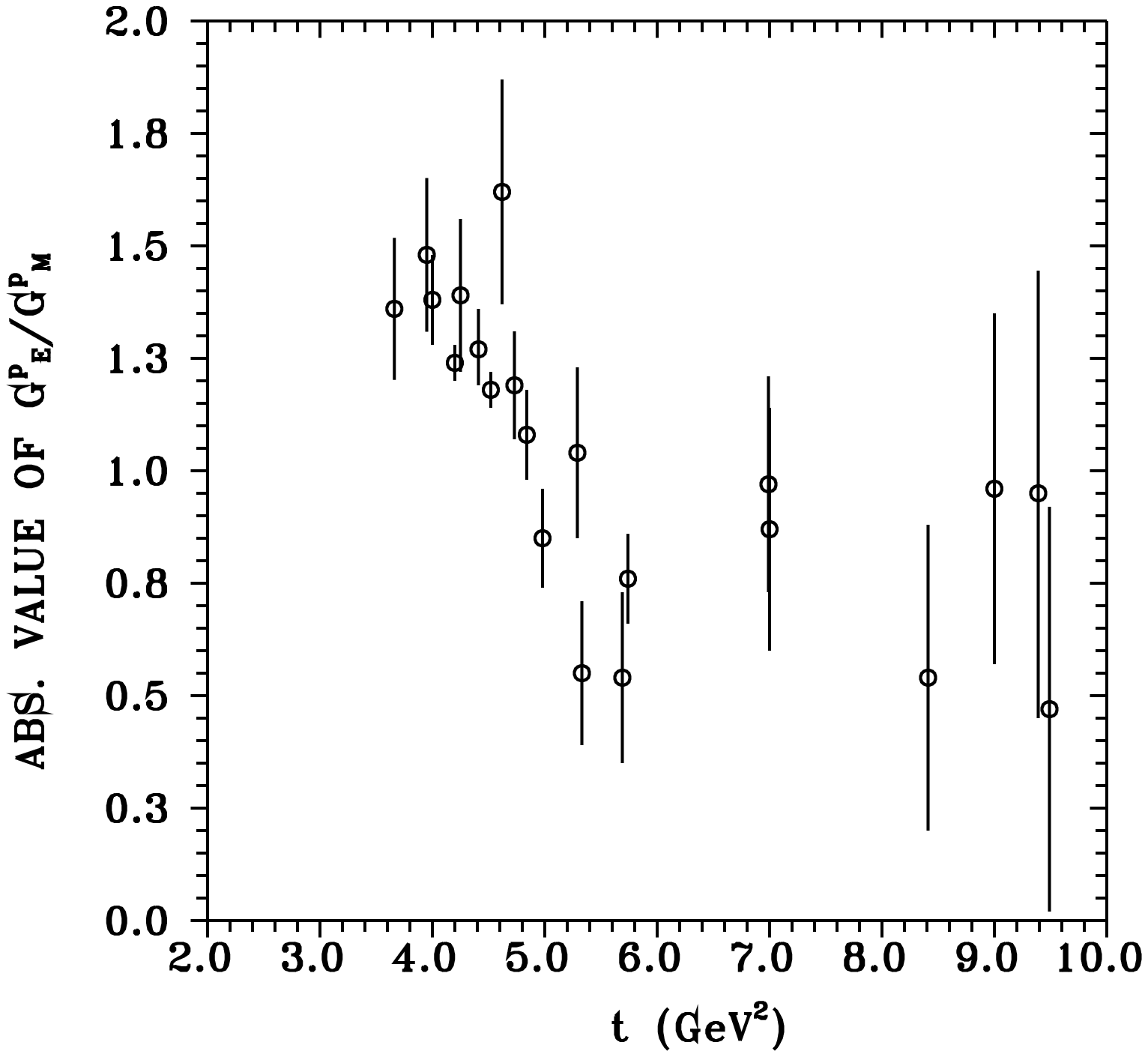}\\
\caption{Experimental data on the ratios of the proton electric to magnetic FFs in space-like and time-like
regions.\label{rpemstd}}
\end{figure}

\begin{figure}
    \includegraphics[width=0.25\textwidth]{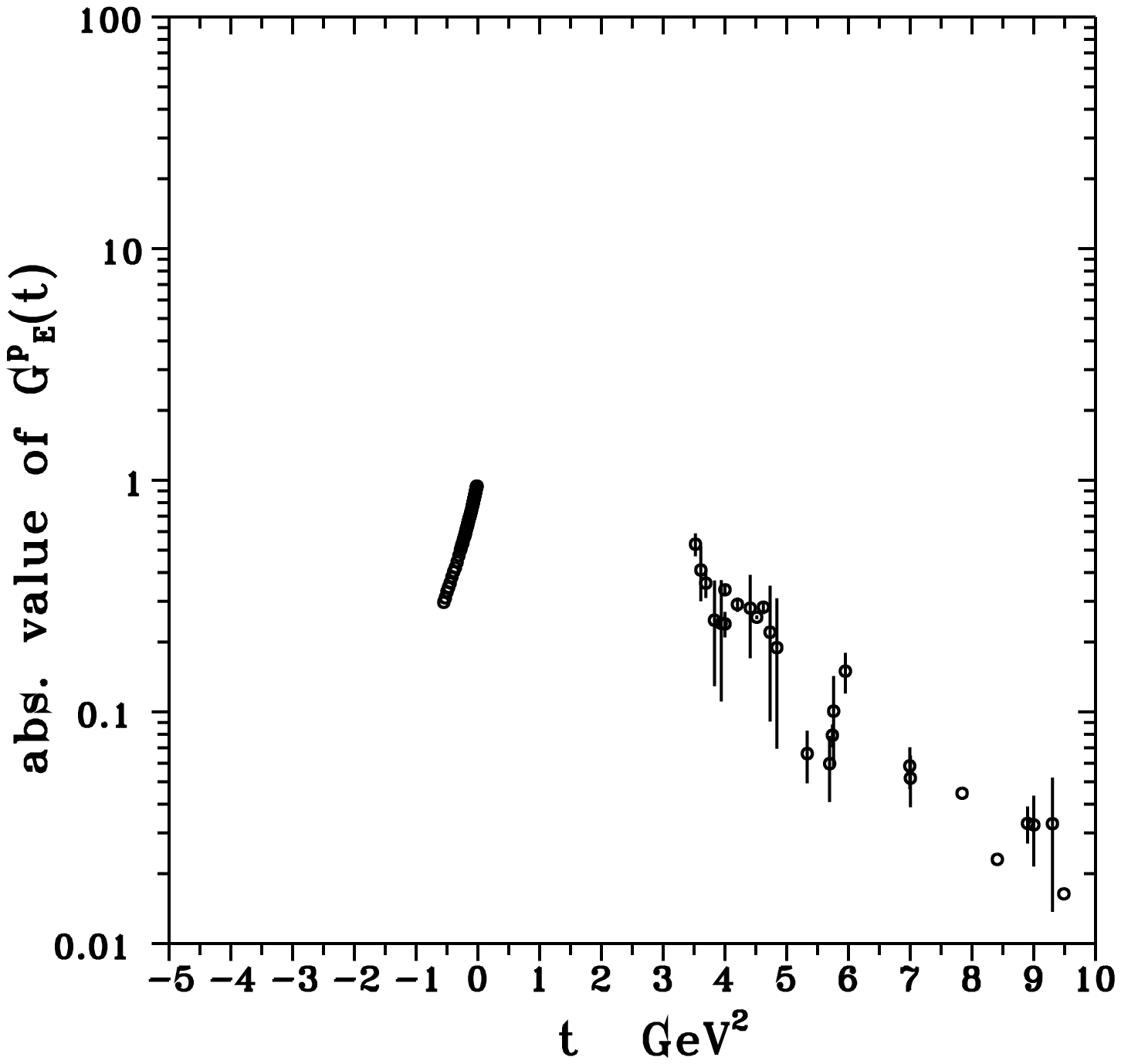}\hspace{0.3cm}
    \includegraphics[width=0.25\textwidth]{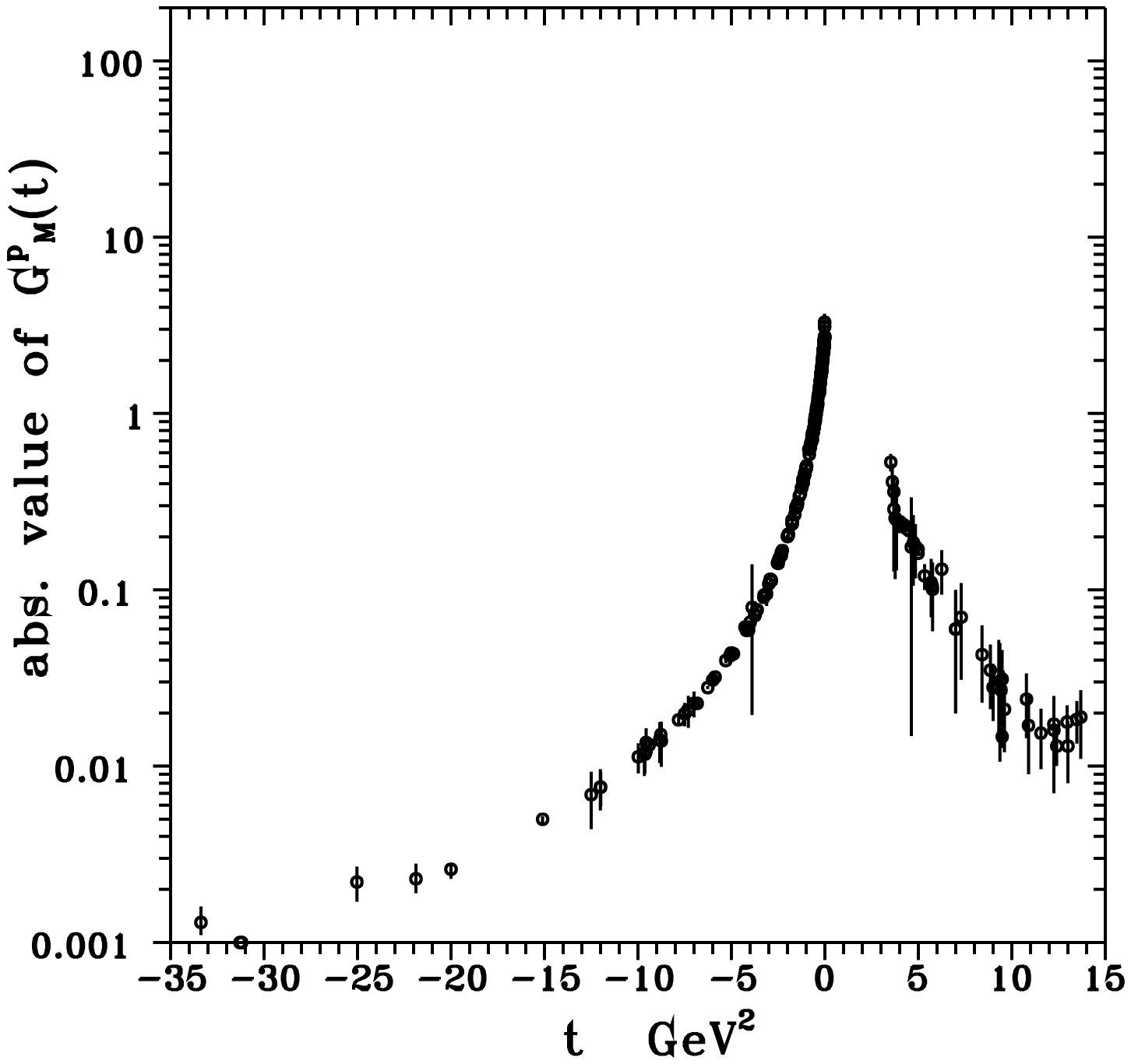}\\
\caption{Experimental data on the proton electric and magnetic FFs in space-like and time-like regions.\label{pemstd}}
\end{figure}

\begin{figure}
    \includegraphics[width=0.25\textwidth]{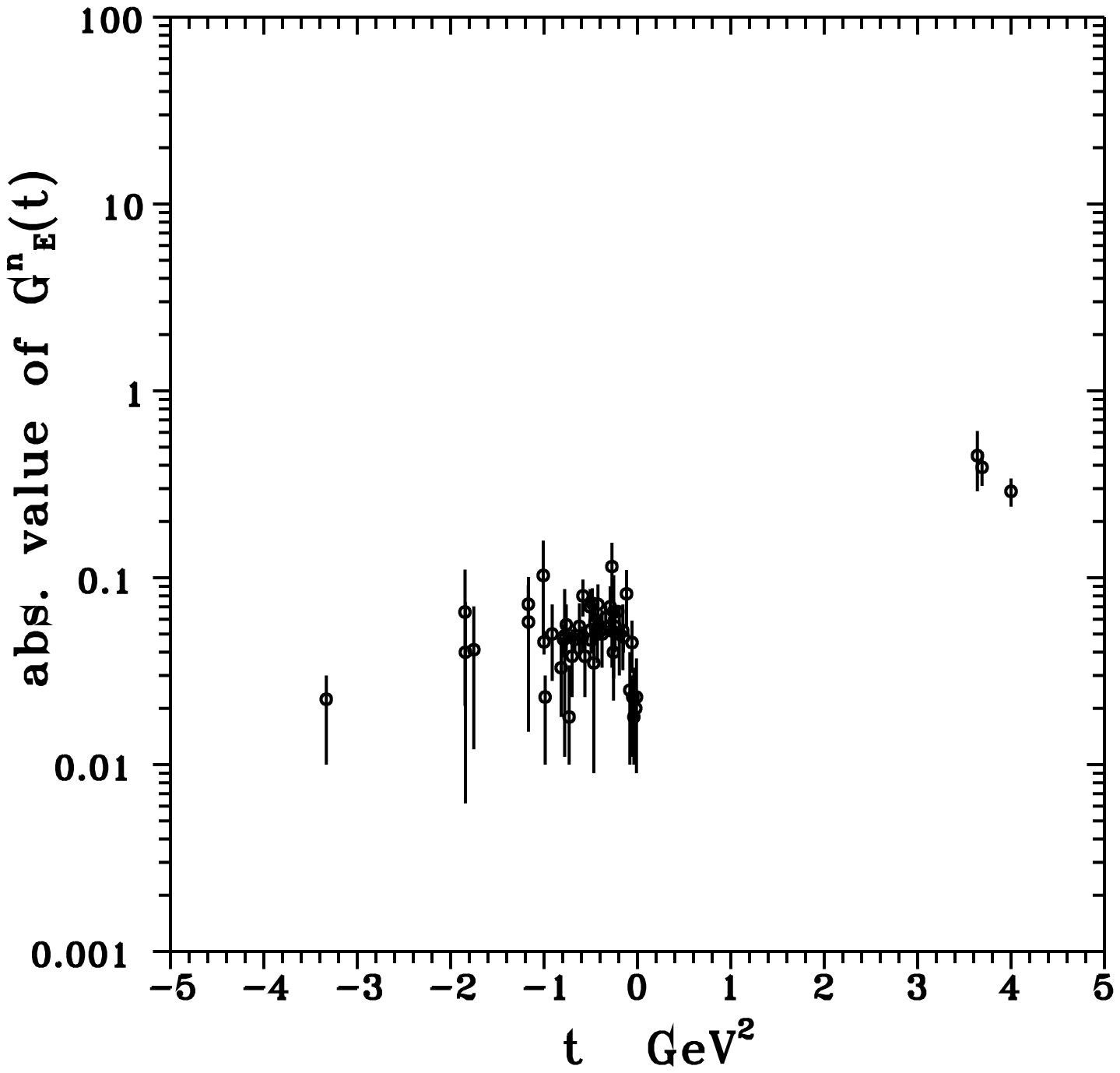}\hspace{0.3cm}
    \includegraphics[width=0.25\textwidth]{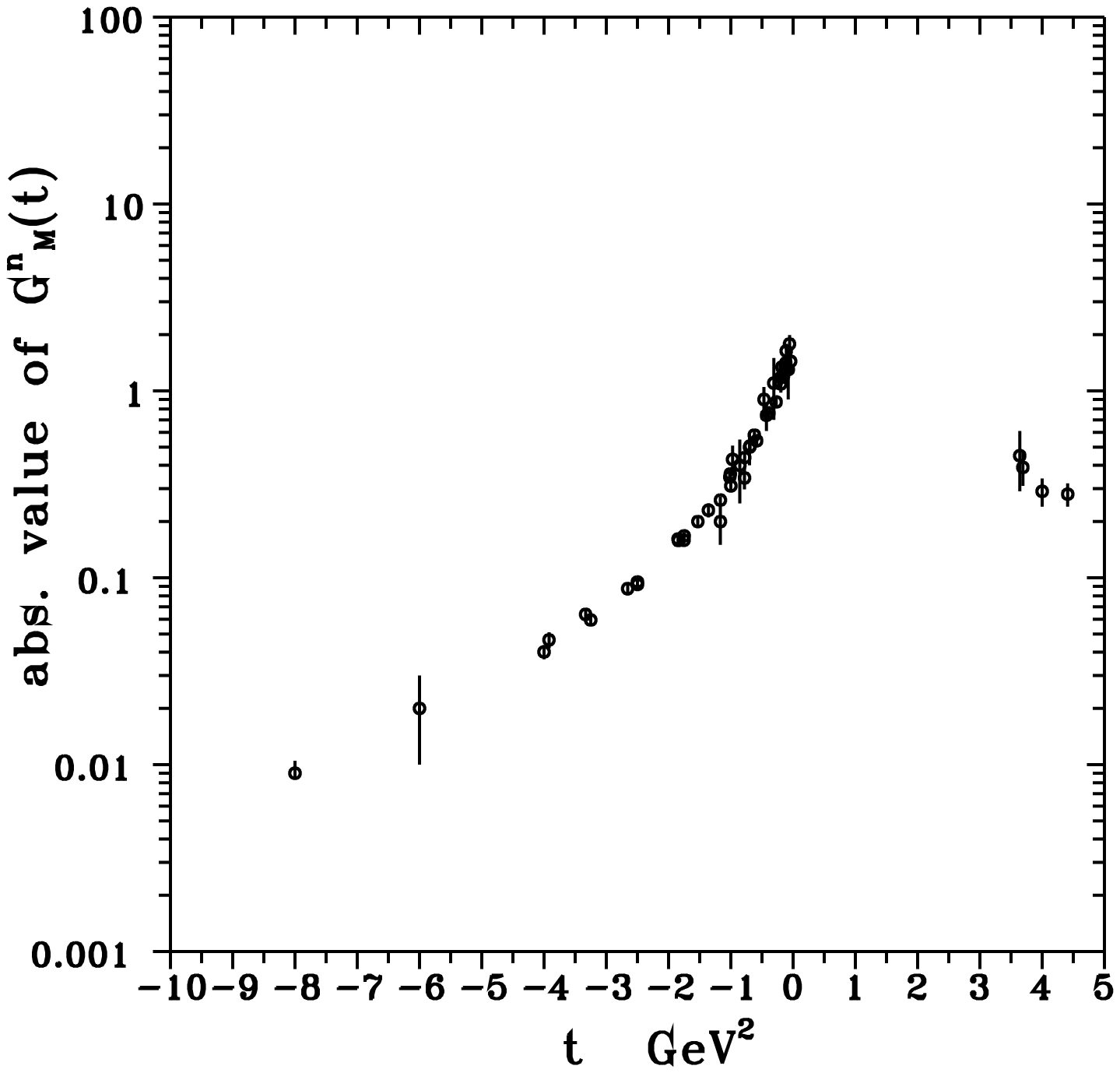}\\
\caption{Experimental data on the neutron electric and magnetic FFs in space-like and time-like regions.\label{nemstd}}
\end{figure}

\begin{figure}
\centering
    \includegraphics[width=0.25\textwidth]{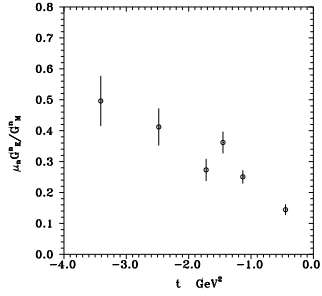}
\caption{Data on the ratio $\mu_nG^n_E(t)/G^n_M(t)$ in the
space-like $(t<0)$ region from polarization experiments on the
light nuclei.\label{rnemsld}}
\end{figure}

   Some of them are mutually contradicting e.g. the proton electric FF data in the spacelike region to be obtained by the Rosenbluth method from measured
$d\sigma/d\Omega (e^-p \to e^-p)$ with data on the ratio $\mu_p G^p_E(t)/G^p_M(t)$ at the same region obtained in the polarization experiments. Therefore further we have striven for a choice of only the most reliable data in the analysis. All these data are graphically presented in Figs. \ref{rpemstd}-\ref{rnemsld}.

   Results of a simultaneous analysis of these more or less 582 reliable experimental points on
$G^N_E(t), G^N_M(t)$ $N=p, n$ and their ratios by the nucleon electromagnetic structure $Unitary\&Analytic$ model to be given by the relations (\ref{FN1s})-(\ref{FN2v})
through (\ref{pEMFFs}) and (\ref{nEMFFs}) with 12 free parameters are given in TABLE I

\bigskip
TABLE I: Results of the analysis of the proton and neutron EM FFs data together\\
  $\chi^2/ndf=1.76;$ \quad $t^{1s}_{in}= (1.4653\pm 0.0542) GeV^2; t^{1v}_{in}= (2.9631\pm 0.0072) GeV^2;$\\
  $t^{2s}_{in}= (1.8513\pm 0.0049) GeV^2; t^{2v}_{in}= (2.3927\pm 0.0039) GeV^2;$\\
  $(f^{(1)}_{\omega' NN}/f_{\omega'})= -0.2780\pm 0.0056; (f^{(1)}_{\phi' NN}/f_{\phi'})= -0.5214\pm 0.0030;\\
   (f^{(1)}_{\omega NN}/f_{\omega})= 0.5988\pm 0.0014; (f^{(1)}_{\phi NN}/f_{\phi})= -0.0287\pm 0.0009;\\
   (f^{(2)}_{\phi' NN}/f_{\phi'})= 0.0422\pm 0.0156; (f^{(2)}_{\omega NN}/f_{\omega})= -0.4872\pm 0.0828;\\
   (f^{(2)}_{\phi NN}/f_{\phi})= 0.1216\pm 0.0032; (f^{(1)}_{\rho NN}/f_{\rho})= -0.0602\pm 0.0026;$\\

\begin{figure}
    \includegraphics[width=0.25\textwidth]{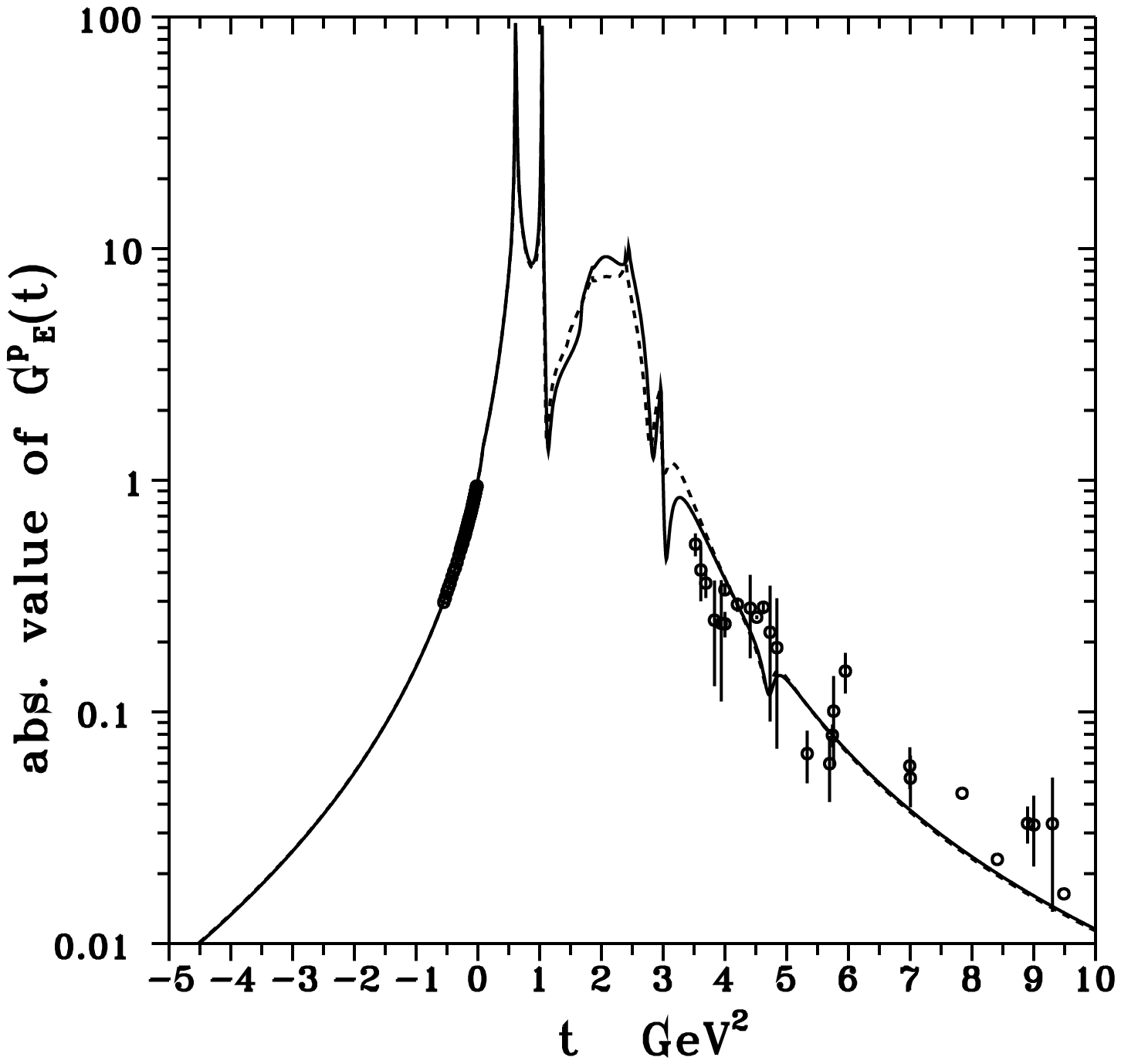}\hspace{0.3cm}
    \includegraphics[width=0.25\textwidth]{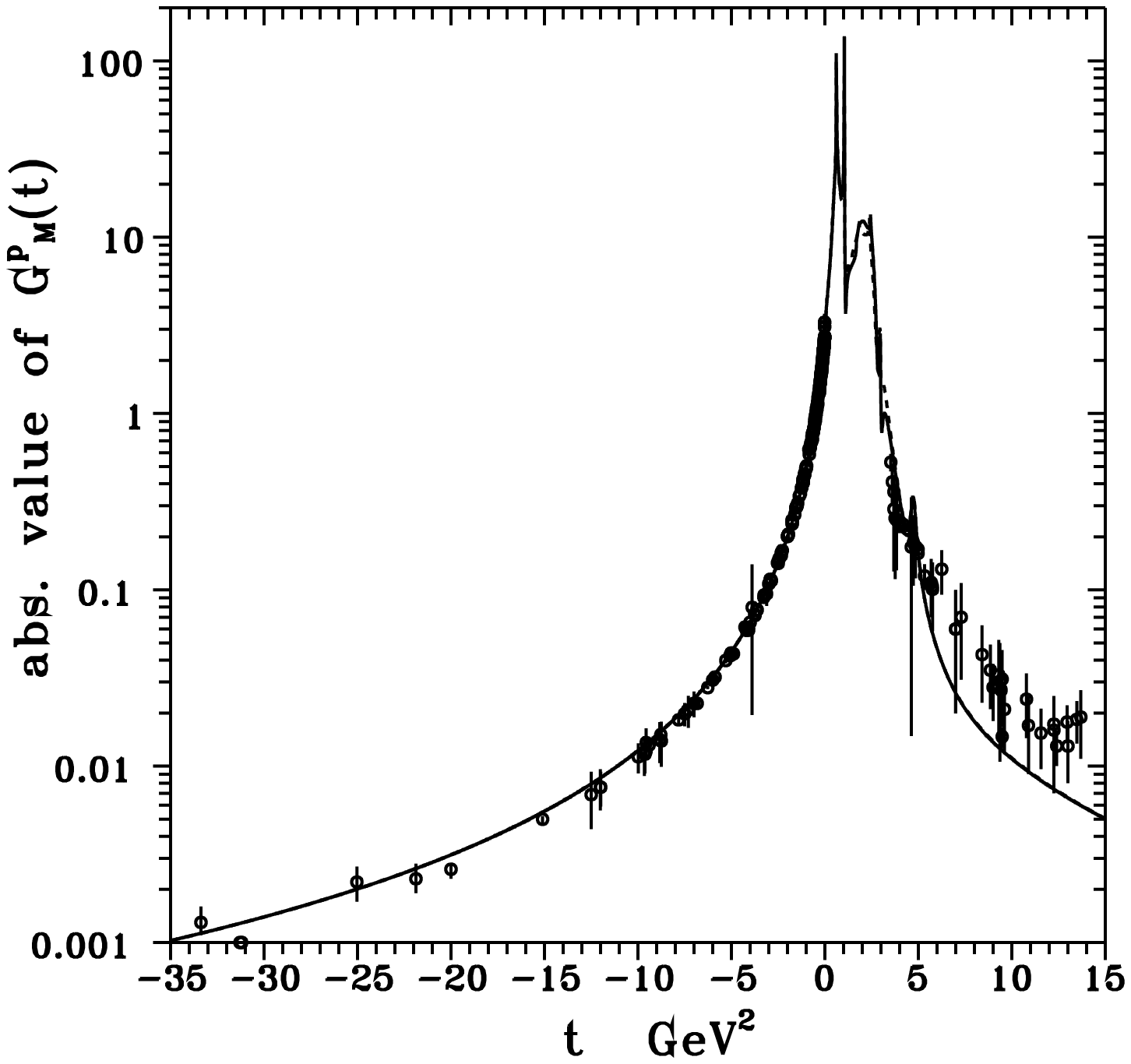}\\
\caption{Prediction of proton electric and magnetic FFs behavior by the nucleon $U\&A$ model \cite {ABDD} and its comparison with existing data.\label{pemstth}}
\end{figure}

\begin{figure}
    \includegraphics[width=0.25\textwidth]{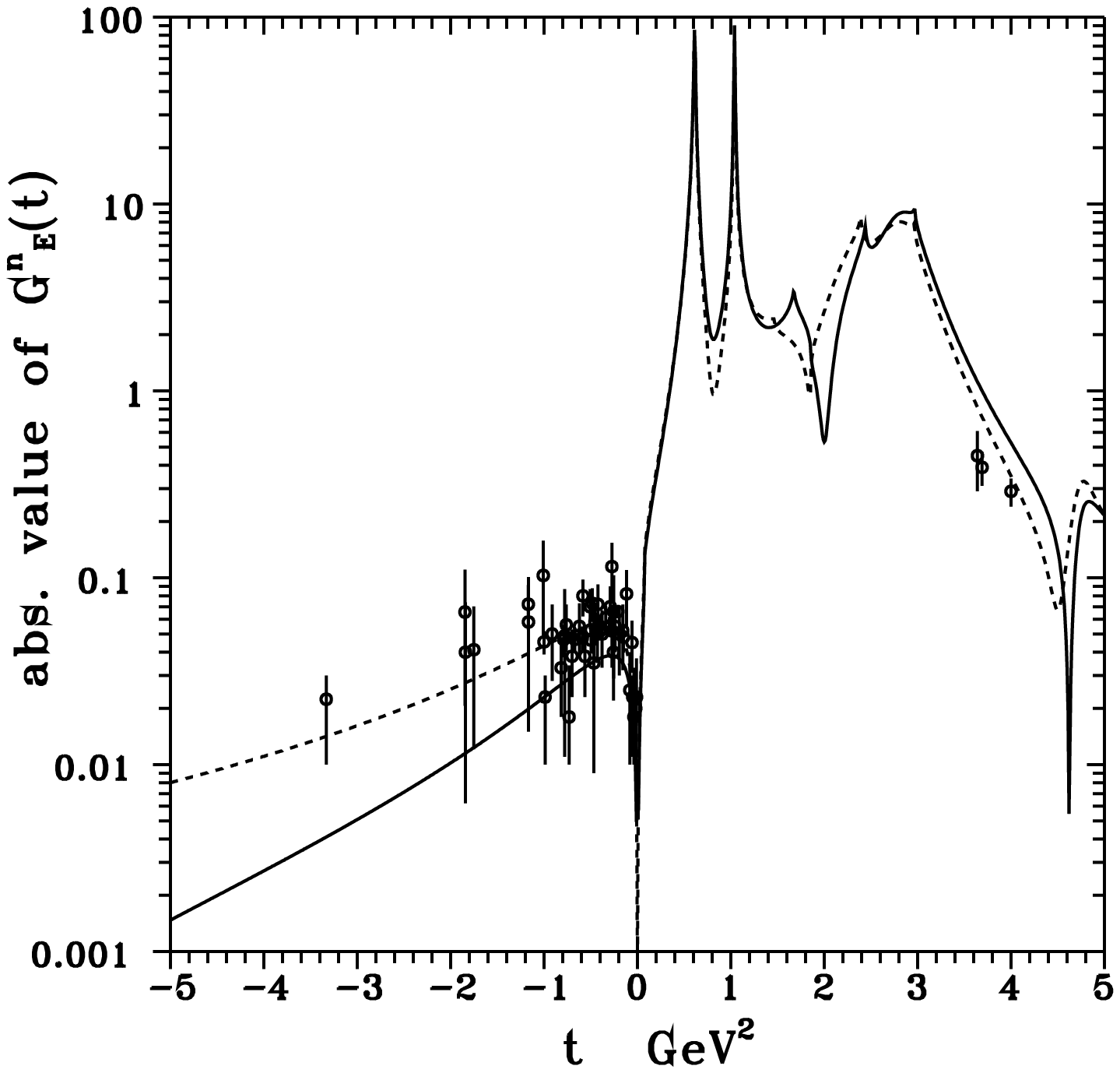}\hspace{0.3cm}
    \includegraphics[width=0.25\textwidth]{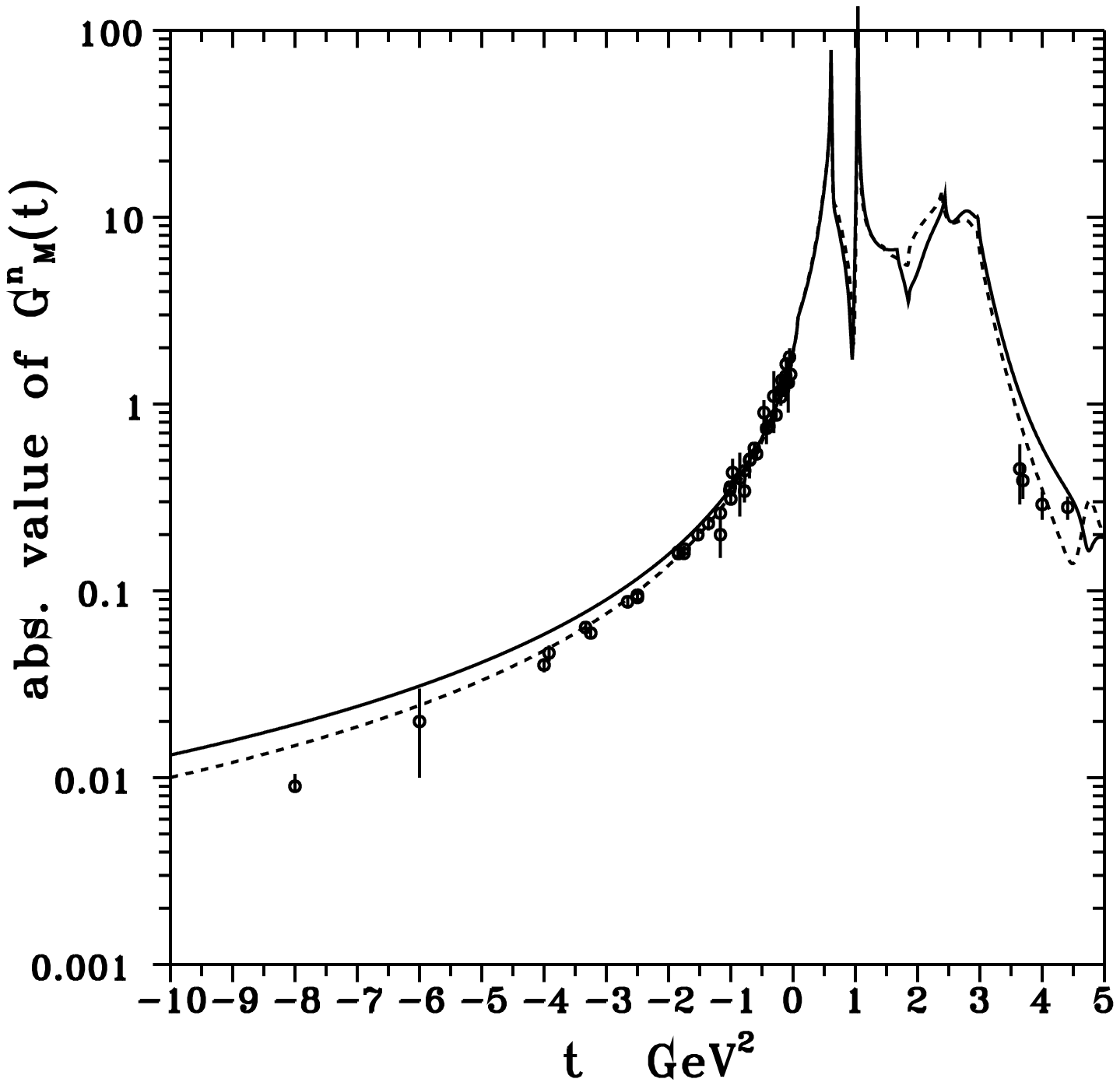}\\
\caption{Prediction of neutron electric and magnetic FFs behavior by the nucleon $U\&A$ model \cite {ABDD} and its comparison with existing data.\label{nemstth}}
\end{figure}

\bigskip
   The corresponding behaviors of the proton and neutron electric and magnetic FFs on the base of these results and their comparison with experimental data are presented
in Figs. \ref{pemstth} and \ref{nemstth} by the dashed lines.

\bigskip
\begin{figure}
\centering
    \includegraphics[width=0.25\textwidth]{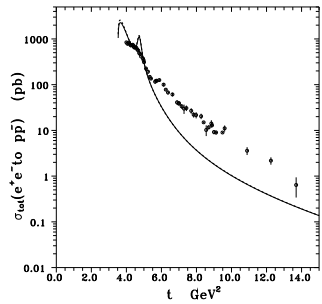}
\caption{Prediction of $\sigma_{tot}(e^+e^- \to p\bar p)$ behavior by
       $G^p_E(t), G^p_M(t)$ from Fig.\ref{pemstth}  and its
       comparison with the recent data \cite{Ab1, Ab2} measured
       at the BEPcII collider.\label{toteepp}}
\end{figure}

   Substitution of these results  on $G^p_E(t), G^p_M(t)$ given by dashed lines in Fig. \ref{pemstth} into relation (\ref{totcspp}) to be modified for protons gives
the theoretically predicted dashed curve in Fig. \ref{toteepp}, where it is compared with very precise data \cite{Ab1, Ab2} on $\sigma_{tot}(e^+e^- \to p\bar p)$
measured by BES III Collaboration at the BEPcII collider exploiting the initial state radiation technique with an undetected photon.

   The Fig. \ref{toteepp} reveals a disagreement between the theoretically predicted $\sigma_{tot}(e^+e^- \to p\bar p)$ by using behaviors of
$G^p_E(t), G^p_M(t)$ from Fig. \ref{pemstth} and the experimental data on $\sigma_{tot}(e^+e^- \to p\bar p)$ \cite{Ab1, Ab2}
recently measured at the BEPcII collider.

   There was our conjecture that the latter result could be caused by the neutron EM FFs data in the analysis as they are considerably less precise
in comparison with the proton data.

   In order to confirm or disprove our hypothesis we have excluded all neutron EM FFs data from the complete nucleon EM FFs compilation and the determination of the
free parameters of the proton electromagnetic structure $Unitary\&Analytic$ model to be given by the relations (\ref{FN1s})-(\ref{FN2v}) and (\ref{pEMFFs}) has been
carried out by the analysis of only 459 reliable experimental data points with errors on the proton EM FFs.

   The results from such analysis are presented in TABLE II

\bigskip
TABLE II: Results of the analysis of only proton EM FFs data\\
  $\chi^2/ndf=1.74;$ \quad $t^{1s}_{in}= (1.6750\pm 0.0363) GeV^2; t^{1v}_{in}= (2.9683\pm 0.0091) GeV^2;$\\
  $t^{2s}_{in}= (1.8590\pm 0.0023) GeV^2; t^{2v}_{in}= (2.4425\pm 0.0208) GeV^2;$\\
  $(f^{(1)}_{\omega' NN}/f_{\omega'})= -0.2937\pm 0.0015; (f^{(1)}_{\phi' NN}/f_{\phi'})= -0.5298\pm 0.0027;\\
   (f^{(1)}_{\omega NN}/f_{\omega})= 0.6384\pm 0.0025; (f^{(1)}_{\phi NN}/f_{\phi})= -0.0271\pm 0.0005;\\
   (f^{(2)}_{\phi' NN}/f_{\phi'})= 0.3075\pm 0.0156; (f^{(2)}_{\omega NN}/f_{\omega})= 0.1676\pm 0.0377;\\
   (f^{(2)}_{\phi NN}/f_{\phi})= 0.1226\pm 0.0035; (f^{(1)}_{\rho NN}/f_{\rho})= -0.0802\pm 0.0014;$\\

\bigskip
   The corresponding behaviors of the proton electric and magnetic FFs on the base of the parameters from
TABLE II are presented in Fig. (\ref{pemstth}) by full lines and their substitution into relation (\ref{totcspp}) to be modified for protons is given by full line
in Fig. \ref{toteepp}. The covering of the previous dashed line by full line confirms our guess about the neutron EM FFs to be unreliable,
which is confirmed also by full lines in Fig. (\ref{nemstth}) for neutron EM FFs to be predicted by the parameters of the TABLE II.

   The latter demonstrates a definite disagreement between the theoretical prediction of
$\sigma_{tot}(e^+e^- \to p\bar p)$ by both $G^p_E(t), G^p_M(t)$ from Fig. \ref{pemstth} represented by full and dashed lines and the experimental data on $\sigma_{tot}(e^+e^- \to p\bar p)$ \cite{Ab1, Ab2}
recently measured at the BEPcII collider.

\section{Conclusions}

   All existing data on the proton and neutron electromagnetic form factors, also on their ratios, have been collected and then by means of the
advanced nucleon electromagnetic structure $Unitary\&Analytic$ model described with the aim of an investigation of their consistency with new
precise data on the total cross section of the electron-positron annihilation into proton-antiproton pairs. A disagreement between the theoretical prediction of $\sigma_{tot}(e^+e^- \to p\bar p)$ and its recently measured data by BESSIII Collaboration, first by using behaviors of $G^p_E(t), G^p_M(t)$ obtained in the
simultaneous analysis of data on proton and neutron EM FFs together, then this result is definitely confirmed also by using behaviors of $G^p_E(t), G^p_M(t)$ obtained in the analysis without the neutron EM FFs data.

\medskip

   The authors would like to thank Eric Bartos for valuable discussions.

\end{document}